\documentstyle[12pt]{article}
\def\Journal#1#2#3#4{{#1} {\bf #2}, #3 (#4)}

\def\MOD{{\em Mod. Phys. Lett} A}

\def\NPB{{\em Nucl. Phys.} B}
\def\PLB{{\em Phys. Lett.}  B}
\def\PRL{\em Phys. Rev. Lett.}
\def\PRD{{\em Phys. Rev.} D}

\def\dalemb#1#2{{\vbox{\hrule height .#2pt
        \hbox{\vrule width.#2pt height#1pt \kern#1pt
                \vrule width.#2pt}
        \hrule height.#2pt}}}

\let\a=\alpha \let\b=\beta \let\g=\gamma \let\d=\delta \let\e=\epsilon
\let\z=\zeta  \let\q=\theta \let\i=\iota \let\k=\kappa
\let\l=\lambda \let\m=\mu \let\n=\nu \let\x=\xi \let\p=\pi \let\r=\rho
\let\s=\sigma \let\t=\tau \let\u=\upsilon \let\f=\phi \let\c=\chi 
\let\w=\omega      \let\G=\Gamma \let\D=\Delta \let\Q=\Theta \let\L=\Lambda
\let\X=\Xi \let\P=\Pi  \let\U=\Upsilon \let\F=\Phi

 \def\bd{\begin{document}} \def\ed{\end{document}}
\def\ds{\documentstyle} \let\fr=\frac \let\bl=\bigl \let\br=\bigr
\let\Br=\Bigr \let\Bl=\Bigl 
\let\bm=\bibitem
\let\na=\nabla
\let\pa=\partial \let\ov=\overline
\def\ie{{\it i.e.\ }} 
\newcommand{\pr}{\paragraph{}}
\newcommand{\be}{\begin{equation}}
\newcommand{\ee}{\end{equation}}
\newcommand{\beba}{\begin{equation}\begin{array}{lcl}}
\newcommand{\eaee}{\end{array}\end{equation}}
\newcommand{\bea}{\begin{eqnarray}}
\newcommand{\eea}{\end{eqnarray}}
\newcommand{\ba}{\begin{array}}
\newcommand{\ea}{\end{array}}
\newcommand{\td}{\tilde}
\newcommand{\norsl}{\normalsize\sl}
\newcommand{\ns}{\normalsize}
\newcommand{\refs}[1]{(\ref{#1})}
\def\simlt{\mathrel{\lower2.5pt\vbox{\lineskip=0pt\baselineskip=0pt
           \hbox{$<$}\hbox{$\sim$}}}}
\def\simgt{\mathrel{\lower2.5pt\vbox{\lineskip=0pt\baselineskip=0pt
           \hbox{$>$}\hbox{$\sim$}}}}
\def\bal{{\mbox{\boldmath $\alpha$}}}
\def\bla{{\mbox{\boldmath $\lambda$}}}
\def\bbe{{\mbox{\boldmath $\beta$}}}
\def\bt{{\mbox{\boldmath $\tau$}}}
\def\bq{{\bf q}}
\def\bd{{\bf d}}
\def\bk{{\bf k}}
\def\bc{{\bf c}}
\def\bw{{\bf w}}
\def\bH{{\bf H}}
\def\bk{{\bf k}}
\def\bx{{\bf x}}
\def\boe{{\bf e}}
\def\a{\alpha}
\def\b{\beta}
\def\g{\gamma}
\def\c{\chi}
\def\d{\delta}
\def\e{\epsilon}
\def\ep{\varepsilon}
\def\FO{\phi}
\def\i{\iota}
\def\z{\psi}
\def\zb{\overline{\psi}}
\def\zt{\widetilde{\psi}}
\def\k{\kappa}
\def\l{\lambda}
\def\m{\mu}
\def\n{\nu}
\def\o{\omega}
\def\p{\pi}
\def\q{\theta}
\def\th{\theta}
\def\tc{\hat{\theta}}
\def\r{\rho}
\def\s{\sigma}
\def\st{\widetilde{\sigma}}
\def\sut{\utw{\sigma}}
\def\t{\tau}
\def\u{\upsilon}
\def\x{\xi}
\def\z{\zeta}
\def\w{\wedge}
\def\D{\Delta}
\def\F{\Phi}
\def\G{\Gamma}
\def\J{\Psi}
\def\L{\Lambda}
\def\O{\Omega}
\def\P{\Pi}
\def\Q{\Theta}
\def\U{\Upsilon}
\def\X{\Xi}
\def\f{\Phi_0}
\def\wtd{\widetilde}
\def\XH{\hat{X}}
\def\DH{\hat{D}}
\def\gh{\hat{g}}
\def\bh{\hat{b}}
\def\sg{\sqrt{-\g}}
\def\pa{\partial}
\def\gh{\hat{g}}
\def\bh{\hat{b}}
\def\mb{{\bar{m}}}
\def\nb{{\bar{n}}}
\def\rb{{\bar{r}}}
\def\sb{{\bar{s}}}
\def\ght{\hat{g}\kern-0.6em \widetilde{\raisebox{-0.12em}{\phantom{X}}}}
\def\bht{\hat{b}\kern-0.6em \widetilde{\raisebox{0.15em}{\phantom{X}}}}
\def\cF{{\cal F}}
\def\bG{{\bf G}}
\def\cL{{\cal L}}
\def\cM{{\cal M}}
\def\cG{{\cal G}}
\def\vf{\varphi}
\def\ft#1#2{{\textstyle{{\scriptstyle #1}\over {\scriptstyle #2}}}}
\def\fft#1#2{{#1 \over #2}}
\def\del{\partial}
\def\sst#1{{\scriptscriptstyle #1}}
\def\oneone{\rlap 1\mkern4mu{\rm l}}
\def\e7{E_{7(+7)}}
\def\td{\tilde}
\def\wtd{\widetilde}
\def\im{{\rm i}}
\def\bog{Bogomol'nyi\ }
\newcommand{\ho}[1]{$\, ^{#1}$}
\newcommand{\hoch}[1]{$\, ^{#1}$}
\newcommand{\ra}{\rightarrow}
\newcommand{\lra}{\longrightarrow}
\newcommand{\Lra}{\Leftrightarrow}
\newcommand{\ap}{\alpha^\prime}
\newcommand{\bp}{\tilde \beta^\prime}
\newcommand{\tr}{{\rm tr} }
\newcommand{\Tr}{{\rm Tr} } 

\begin{document}
\thispagestyle{empty}
\rightline{CERN-TH/98-317}
\rightline{hep-ph/9809582}
\vspace{1truecm}

\centerline{\bf \Large 
Phenomenology 
of Low Quantum Gravity Scale Models}

\vspace{1.2truecm}
\centerline{{\bf Karim Benakli} }
\vspace{.5truecm}
%{\em

\centerline{CERN, Theory Division,
CH-1211 Geneva 23, SWITZERLAND}

\vspace{2.2truecm}

%%%%%%%%%%%%%%%%%%%%%%%%%%%%%%%%%%%%%%%%%%%%%%%%%%%%%%%%

\vspace{.5truecm}

\begin{abstract}

We study some phenomenological implications of models where the 
scale of quantum gravity effects lies much below the  four-dimensional 
Planck scale. These models arise from M-theory vacua where   
either the internal space volume is large   or  
the string  coupling is very small. We provide  
a critical analysis of ways to unify electroweak, strong
and gravitational interactions in M-theory.   We  
discuss the relations between different scales in two  M-vacua: 
Type I strings and  Ho\v rava--Witten supergravity models. The latter 
allows possibilities for an eleven-dimensional scale at  TeV energies
with {\em one} large dimension below  separating our 
four-dimensional world from a hidden one. Different  
mechanisms for breaking supersymmetry (gravity mediated, gauge mediated and 
Scherk-Schwarz mechanisms) are discussed in this framework. Some 
phenomenological issues such as dark matter (with
masses  that may  vary in time), origin of neutrino masses and axion scale are 
discussed. 
We suggest that these are indications that
the string scale may be lying in the $10^{10}$--$10^{14}$ GeV region.

\end{abstract}

CERN-TH/98-317

September 1998

\newpage
%%%%%%%%%%%%%%%%%%%%%%%%%%%%%%%%%%%%%%%%%%%%%%%%%%%%%%%%%%%%%%%%

\section{Introduction}

One of the fundamental questions of particle physics is  about the
ultimate structure of particles like quarks and leptons.  It is
believed that when probing shorter distances one would reach scales
where  quantum  gravitational effects become important. As gravity
seems to deal with geometry, these effects  may just render invalid
our basic notions as shapes and length used to study macroscopic
objects.  M-theory is supposed to provide us with the formalism
necessary to study and formulate the laws governing  physics at such
small distances. There the fundamental objects of M-geometry are no
more points but  p-dimensional extended objects: p-branes.

A crucial question is then: At which scales $M_s$ do quantum
gravitational effects become important?.  Simple dimensional analysis
of the low energy parameters  lead to a value of the order  of
$M_s^{-1} \sim M_{P}^{-1} \sim 10^{-33}$ cm. However the structure of
space-time might change at much bigger length scale  leading to
changements of the strength of gravitational interactions for instance
in which case $M_s$ can be much lower. The existence of vacua of
M-theory which would allow to decrease this scale has been pointed out
by Witten \cite{witten}. He suggested that $M_s$ could correspond to
scales of the order of $10^{16}$ GeV where the three known gauge
interactions have been argued to unify \cite{LEP1, LEP2} in the simplest
supersymmetric extension of the standard model: the Minimal
Supersymmetric Standard Model (MSSM).

The scale $M_s$ may in fact lie at much lower values. Experimental
bounds on the effects of excitations of standard model particles as
higher order effective operators \cite{AB} and form factors in the
gauge interactions \cite{caceres} exclude only the region with $M_s$
less than few TeV. That $M_s$ lies just above the electroweak scale
was proposed  by a number of authors\footnote{ The possibility that
part of the string spectrum corresponding to the Kaluza-Klein
excitation of one or two large dimensions lie at the TeV scale  was
suggested earlier in \cite{Ant}.}
\cite{lykken,bachas1,ADD,AADD}. In particular a viable scenario
based on some early field theory analysis in \cite{ADD} was exhibited 
in \cite{AADD}. Possible realization of such a
scenario in Type I strings has  been investigated in \cite{AADD,
tye}. It implies that future accelerators might be able to discover
the existence of extra-dimensions \cite{AADD,ABQ} and string-like
structure of matter \cite{AADD}.

The precise mechanism of unification of coupling constants in these
scenarios is an important issue. The scale $M_s$ and the size of
internal dimensions are closely related to  the strength of the
couplings.  The relations between these entities are usually known at
$M_s$. Tthey involve the values of coupling at much higher energy
scales than those where measurements are performed.  Relating these
two values is then necessary before computing the scales. The
possibility of using large thresholds to achieve unification has been
implemented recently in the case of low $M_s$ by \cite{Dudas} and
\cite{bachas2}. Differences between the two works due to the use of an
heterotic string cut-off in the first and a Type I cut-off for the
second illustrates  the fact that such thresholds have to be computed
in a full M-theory framework. Here we will propose that unification
might happen naturally in even simpler ways. For instance simple
models might unify at intermediary regions, after logarithmic running,
either through conventional or rational unification. We discuss these
issues in Section 2.

Once the coupling constants are known, the size of the other
parameters of the theory can be computed as to fit the observed value
for the strength of  gravitational interactions. The latter are known
to be very weak at low energies. From the point of view of M-theory
this can be due to different reasons: (a) the scale $M_s$ which
suppress them is very large, (b) $M_s$ is low but as the internal
space is large, (c) the coupling constant is extremely small at the
string scale and gauge couplings grow rapidly below $M_s$ while the
gravitational coupling either grows slowly or remains constant. The
case (a) is the conventional one.  The case (b) has attracted recently
most of the attention \cite{ADD}. While the case (c) of which a version was
proposed in \cite{bachas1} has not been discussed further. The main
problem with such a scenario is that one appeals to very large
thresholds to drive the gauge couplings from nearly vanishing to order
one values in order to comply with the observations. Computations of
such thresholds have to be done in a fully M-theoretical framework and
such models are not yet available. However this scenario is worth
studying as it illustrates the possibility that quantum gravitational
effects are never big.

There are two classes of M-vacua that are simple and suitable to
discuss Low Quantum Gravity Scale (LQGS) models.  The first one is
Type I strings \cite{Irev}. These allow a stringy realization of the proposal 
of \cite{ADD} and they offer the advantage that full M-theory
computations may be carried on.  Another class is the M-theory on
$S^1/Z_2$ of Ho\v rava--Witten \cite{HW}. In both cases, the low
energy picture is  of worlds living on three-branes separated by a
bulk where gravitons propagate  \footnote{ Attempts to describe  Ho\v
rava-Witten models as branes might  be found in \cite{stelle}.}. All
precedent authors claimed that the lowest possible value for the
eleven dimensional Planck mass $M_{11}$ is around $10^7$ GeV and thus
Ho\v rava--Witten compactifications are excluded for TeV-LQGS models. However, 
these results were deduced with the assumption was made
that the six-dimensional Calabi-Yau volumes which determine the gauge and gravitational strength
are of the same order. As it was shown in \cite{nse} (see also
\cite{POK}) the average volume which determines the Newton constant
may be much bigger than the volume on the observable wall. This allows
to have scales  $M_{11}$ (hence eleven dimensional physics) at energies as 
small as few TeV!. We discuss these issues
in Section 3.

If the scale $M_s$ lies much above the TeV region then one may suppose
that the theory is  supersymmetric at higher energies
i.e. supersymmetry is broken in our observable world at scales around
the TeV.  The most popular mechanisms to achieve the supersymmetry
breaking may be put in three categories. The first assumes  gravity
mediates supersymmetry breaking from a hidden sector to the observable
one \cite{gc, Mgc}.  The hidden sector might either be on our wall, in
the bulk or on another wall. The second scenario assumes that
supersymmetry breaking is mediated through gauge interactions
\cite{gm,Peskin}.  In the latter two cases Kaluza-Klein states
contribute to the mediation of supersymmetry breaking. The same
remarks hold  in the case of gauge mediation \cite{gm,Peskin}. Another
possibility is to use the Scherk-Schwarz mechanism  where
supersymmetry is spontaneously broken at tree level by non-trivial
periodicity condition for supersymmetric partners different in some
compact  internal dimension \cite{ss,ssf}.  We discuss theses issues
in  Section 4 and comment on effects on soft-masses.

In section 6 we  discuss some possibilities to have dark matter on the
other wall of the universe as suggested for example in the M--theory scenario 
in \cite{BEN,nse} and more specifically within the framework of \cite{ADD}
re by \cite{ADD2}.  We notice that
 these might provide candidates for dark matter with variable
masses. We  comment on neutrino masses and then argue that present
experimental data may be taken as indications that a natural value for
the string scale is $10^{10}$--$10^{14}$ GeV.

In any case the ratio between the Planck mass and the the electro-weak scale 
needs to be explained, probably through some dynamical mechanism that leads to the necessary values for the moduli (radii and couplings) \cite{ADMR}. 
We will not address this issue here.

 Section 7 gives a summaries  our main results.

\section{ Unification of Gauge Couplings in M-theory}

By definition M-theory provides a unified theory for all gauge and
gravitational interactions. This ``unification'' might be achieved in
many ways which contrast with the historic meaning of the word. Below
we will  provide a critical view on  the different possibilities that
have emerged from the study of  several $M$-theory vacua.  In each
case we will discuss the advantages and  shortcomings when applied to
LQGS models. This list might not be exhaustive  as the subject is
still in development.

In its present form, the ``unification'' idea  is an attempt to
explain  the low energy  parameters of the standard model as
predictions of the structure and  dynamics of $M$-theory. In contrast
with early ideas, this does not preclude  the existence of a Grand
Unified Theory (GUT) group where the standard model  symmetry $SU(3)_c
\times SU(2)_w \times U(1)_Y$ is embedded. The  simplest set of
experimental data one may try to ``explain'' are the measured values
at $m_Z=91$ GeV of the strong and electromagnetic couplings $\alpha_s$
and  $\alpha_{em}$ respectively, as well as the value of
$\sin^2{\theta}_w$ where ${\theta}_w$ is the weak angle. The best fit
of different low energy cross sections corresponds to:

\begin{equation}
\alpha_{em}(m_Z)\sim {1\over 128}, \qquad \, \, \, \,
\sin^2{\theta}_w (m_Z) = 0.231 \pm 0.003, \, \qquad \, \, \, \,
\alpha_s (m_Z)  = 0.108 \pm 0.005
\label{expcoup}
\end{equation}

These quantities are related to the gauge couplings of the standard
model group $SU(3)_c \times SU(2)_w \times U(1)_Y$ as  $\alpha_s
\equiv \alpha_3$, $\alpha_{em} = {{\alpha_1 \alpha_2}\over {\alpha_1
+\alpha_2}} $  and $\sin^2{\theta}_w = {{\alpha_1 }\over {\alpha_1
+\alpha_2}}$.

To proceed further,  theoretical inputs which are very much subject to
prejudice, are necessary. For instance one specifies a number of
additional  particles with masses at  intermediate  scales
$M_{in}^{(i)}$  and given gauge  quantum numbers. There are  two
strategies that might be followed. Either one investigates the
existence and the value of possible unification scales for a model (as
MSSM ) with well defined particle content, or one  makes a choice for
$M_s$ then  sets of possible spectra of particles necessary to achieve
the unification are exhibited.

Climbing up the energies, looking for the unification scale, one makes
use of renormalization group equations.  The standard model couplings
at $m_Z$ are related to  the string scale $M_s$ through:

\begin{equation}
\frac {1}{\alpha_a (m_Z)}= \frac {k_a}{\alpha} +  b_i^{(0)} \ln
({M_{s} \over m_Z}) + \Sigma_{n=1}^{N}(b_i^{(n)}- b_i^{(n-1)}) \ {\ln
({M_{s} \over M_{in}^{(n)}})}+\Delta_a
\label{rge}
\end{equation}

where $\Delta_a$ contain higher loops and threshold corrections.

The beta-function coefficients $b_i^{(n)}$ take into account the
contribution of new states that appear at each intermediary scale. For
$M_s \gg $ TeV we assume that the  hierarchy between  gauge different
scales  is stable  because of the presence of low energy
supersymmetry. In this case:

\begin{equation}
b_3^{(0)}=-3 \; \qquad b_2^{(0)}=1\; \qquad b_1^{(0)}=11
\label{eq: nine}
\end{equation}

is a good approximation (at the level of our discussion). For $M_s$
close  to the TeV  region the beta-function coefficients take  their
standard model values:
\begin{equation}
b_3^{0}=-7 \; \qquad b_2^{0}=-\frac {10}{3}\; \qquad b_1^{0}=4
\label{betasm}
\end{equation}

The  parameters $k_a$ in (\ref{rge}) account for different
normalization or different origin  for each of the three couplings.
It is natural to discuss the unification  as function of the allowed
value of $k_1 / k_2$ and $k_2 / k_3$ :

\begin{equation}
{k_1 \over k_2}= { {1-\sin^2{\theta}_w- {\alpha_{em} \over {2 \pi}}
\left[b_1^{(0)} \ {\ln ({M_s \over m_Z})}+ \Sigma_{n=1}^{N}(b_1^{(n)}-
b_1^{(n-1)}) \ {\ln ({M_s \over M_{in}^{(n)}})}+ \Delta_1 \right]}
\over  {\sin^2{\theta}_w- {\alpha_{em} \over {2 \pi}} \left[b_2^{(0)}
\ {\ln ({M_s \over m_Z})} + \Sigma_{n=1}^{N}(b_2^{(n)}- b_2^{(n-1)}) \
{\ln ({M_s \over M_{in}^{(n)}})}+ \Delta_2 \right]}}
\label{k1k2}
\end{equation}

\begin{equation}
{k_2 \over k_3}= {  {{\sin^2{\theta}_w \over \alpha_{em}}- {1 \over {2
\pi}} \left[b_2^{(0)} \  {\ln ({M_s \over m_Z})} +
\Sigma_{n=1}^{N}(b_2^{(n)}- b_2^{(n-1)}) \ {\ln ({M_s \over
M_{in}^{(n)}})} + \Delta_2 \right]} \over  {{1 \over \alpha_s}- { 1
\over {2 \pi}} \left[b_3^{(0)} \  {\ln ({M_s \over m_Z})}+
\Sigma_{n=1}^{N}(b_3^{(n)}- b_3^{(n-1)}) \ {\ln ({M_s \over
M_{in}^{(n)}})} + \Delta_3 \right]}}
\label{k2k3}
\end{equation}
 
The unification scenarios might be considered as of two kinds: those who 
involve only light (four-dimensional) degrees of freedom and those who 
make use of large threshold corrections generated by states propagating 
in the compact internal dimensions.

\subsection{Conventional unification :}

 This scenario assumes that ${k_1 \over k_2}={5\over 3}$ and  ${k_2
\over k_3}=1$ and the threshold corrections $\Delta_a$ are either
small or universal. In the latter case unification is achieved but the 
value of the coupling constant receives sizable  corrections 
from unification to string scale.
Hence, the unification scale is very close to the string scale
$M_s$. It is  a very popular scenario as low energy data \cite{LEP2}
seem to indicate that the MSSM gauge couplings meet at a unification
scale $\sim 10^{16}$ GeV which is close to the Planck scale.  Here we
want to briefly comment on the application of this scenario to the
case of LQGS models.  \vskip 0.5cm $\bullet$ {\em  Mechanism:}

 With the content of the MSSM the unification scale is around  $
10^{16}$ GeV.  To get lower values, additional states  must be present
at intermediairy scales such that   equations (\ref{k1k2}) and
(\ref{k2k3}) lead to   ${k_1 \over k_2}={5\over 3}$ and  ${k_2 \over
k_3}=1$. For any given value  of $M_s$  many  solutions exist. However
very few of these spectra are otherwise motivated.

\vskip 0.5cm $\bullet$ {\em  Size of couplings:}

In the absence of gauge enhancement at intermediate scales,  the
coupling of $U(1)_Y$ increases with  energy. This implies that:

\begin{equation}
 {\alpha} > {k_1}{\alpha_1(M_Z)} \sim \frac {k_1} {100}
\label{bound1}
\end{equation}

In a minimal scheme $k_1 =1$ or 2 and  $1/100 < \alpha < 1$. The upper
value is required in order to keep the perturbative approximation of
the low energy effective theory valid.  

Such a bound might be avoided in  the presence of gauge 
enhancements. In this case the coupling
constants might be pulled logarithmically to lower values. Also 
large threshold corrections (as discussed above) might modify this value.

\vskip 0.5cm $\bullet$ {\em  Advantages:}

The relations ${k_1 \over k_2}={5\over 3}$ and  ${k_2 \over k_3}=1$
allow to embed the  standard  model in a GUT group \cite{ross} as
$SU(4)\times SU(2)^2$ or $SU(3)^3$ (with a discrete $Z_3$
symmetry). This embedding might explain the value of ${k_1 \over k_2}$
that leads (within the appropriate extension of the standard model)
 to $\sin^2{\theta}_w (m_Z)\sim 0.231$ as measured at present colliders.

\vskip 0.5cm $\bullet$ {\em Shortcomings:}

Semi-simple GUT groups structure unify quarks and leptons. This makes
it hard to exhibit   symmetries that would forbid  protons from
decay. So, these groups are not allowed at scales $M_s \ll 10^{16}$
GeV.  Moreover as new particles have to be introduced in the
intermediary region between the electroweak scale and $M_s$ then one
does not predict but fits the values ${k_1 \over k_2}={5\over 3}$ and
${k_2 \over k_3}=1$.

In this scenario, the unification might arise just
below the string scale, at the scale of the compact internal manifold.
In the presence of large ratio of masses (of
Kaluza-Klein states or winding modes compare to the string scale), to 
identify the unification value of the gauge couplings with the string coupling
requires that threshold corrections are small. Nontrivial 
constraints have to be imposed on the string compactification. 
For instance the excitations of standard model
particles might be in (spontaneousely broken) $N=4$ representations \cite{Ant}.

\vskip 0.5cm $\bullet$ {\em Example}
 
Take for instance the left-right extension of MSSM with  an extra pair
of Higgs doublets at a few TeV and a right-handed scale of $10^8$
GeV \cite{DM}. This is a natural candidate to study for low scale 
unification at $M_s\sim 10^{11}$ GeV. Further models will be given elsewhere
\cite{elena}.

\subsection{ Rational unification:}

This possibility has emerged in heterotic string models where  the
parameters $k_a$ are the levels of Kac-Moody algebras on the
world-sheet.  It constitutes a  clear departure from what was
previously referred to as unification. In heterotic string
derived models, proposal to vary $k_1/k_2$ was made in \cite{ibanez},
while to allow also $k_2/k_3$ vary was  proposed in \cite{BS,DFM}.

\vskip 0.5cm $\bullet$ {\em  Mechanism:}

Models with rational unification, i.e. arbitrary $k_a$, can be
constructed the following way\footnote{To classify  ways to  realize
this scenario is an open problem in string theory.}: Consider $k_a$
copies  of a non-abelian group $G_a$ all with the same gauge coupling
constant $g$. An appropriate choice of representations allows to break
spontaneously this symmetry to its
diagonal subgroup.  For example in the case of $k_a=2$ this can be
achieved  by using Higgs fields in bi-fundamental representation. The
result is a   non-abelian factor $G_a$ with gauge coupling
$g/\sqrt{k_a}$. If all the non-abelian  gauge couplings are related to
the same fundamental (string) coupling as  $g_a=g/\sqrt{k_a}$ then we
have achieved rational unification.  In this way the constants $k_a$ have to 
be positive integers for non-abelian groups.

\vskip 0.5cm $\bullet$ {\em  Size of couplings:}

The same arguments used for the case of conventional unification hold
here. 

\vskip 0.5cm $\bullet$ {\em   Advantages:}

This scenario offers  the possibility  of discuss unification without
GUTs at the field theory level for models that would  have otherwise
been thought  non-unified (as  for left-right models in
\cite{BS}). The construction of similar models as described above is
very simple.

\vskip 0.5cm $\bullet$ {\em  Shortcomings:}

From the practical point of view one computes an approximative  real
value for  the ratio of  $k_a$s. This  has to be identified with a
rational number. This is awkward in the absence of  precise estimate
of the higher loops and threshold corrections. If one assumes  the
latter to be negligeable then rational unification requires sometime
large values for  $k_a$s which are not appealing. Moreover the
corresponding string constructions  lead to additional  (undesirable)
light states (but this is a curse on all known string models).

\vskip 0.5cm $\bullet$ {\em   Example:}

Consider extending the MSSM up to energies of order  $\sim 2.5 \times
10^{6}$ GeV  just below $M_s$. Rational unification is obtained at
this scale for  $k_2=1$, $k_3=2$ and $k_1\sim 3$.

\subsection{Accelerated unification:}

In this scenario the thresholds corrections are large and might play an
important role in the  unification process. In the  framework of LQGS
models they have been used in \cite{bachas1} and  \cite{Dudas}.

\vskip 0.5cm $\bullet$ {\em Mechanism:}

The presence of  heavy states charged under the standard model gauge
group leads to threshold corrections.  These might become large if the
typical mass scale of the new states is hierarchically smaller or
bigger than the string scale and their number is large
(infinite). Examples of such states  are Kaluza-Klein excitations,
winding and massive oscillator modes of strings and other $p$-branes.

\vskip 0.5cm $\bullet$ {\em  Size of couplings:}

The size of the coupling as determined by the effective field theory
running  at low energy might be very different from their actual value
at $M_s$. For instance the latter can be  hierarechically smaller or
bigger  if threshold corrections are very large and negative or
positive respectively.

\vskip 0.5cm $\bullet$ {\em  Advantages:}

This scenario allows one  to change the values of the coupling constant
rapidly in a short  energy distance\footnote{ In  \cite{bachas1}  it was suggested
that in heterotic compactifications the Kaluza-Klein excitations might
then drive the coupling constant quickly to  very small values. In
heterotic string compactifications one consequence of such a scenario
is that quantum gravitational quantum effects are weak at the string
scale. In more generic vacua of  $M$-theory  the gauge couplings and
gravitational couplings are functions of different moduli. Both may be
renormalized. Thus at $M_s$ gauge  symmetries may turn to  global
ones, and quantum gravitational may be strong or extremely weak.}. The gauge couplings might then be
driven to unification \cite{Dudas}.

\vskip 0.5cm $\bullet$ {\em Shortcomings:}

The precise scenario depends on the particular M- (string) theory realization 
of this mechanism as it envolves the knowledge of:

1- The threshold corrections  involve an infinite number of states and
thus {\em must} be computed  in a full M-theory framework. For
instance the result of the computation depends on the cut-off which is
different for different string types \cite{BF}.

2- Wilson lines for instance could  introduce very light states
with exotic gauge quantum numbers \cite{WL}.  Changing slightly the
beta-function coefficient can change dramatically the unification process
due to the power law behavior.

3- The spectrum of heavy modes in Calabi-Yau compactifications is
generically difficult to compute. If instead one uses orbifold
compactifications   there are twisted states that are generically
charged under the  standard model  gauge group.  These states
introduce mixings between different KK levels \cite{AB} that could
make the gauge coupling behavior with  energy different from the one
of a purely higher dimensional theory (the theory remembers the
boundaries because  we are  computing  the effects of the
corresponding states).

\vskip 0.5cm $\bullet$ {\em Example:}

Recently an interesting observation was made in \cite{Dudas} that
N=2 supersymmetric multiplets of standard model gauge bosons with or
without  matter might accelerate conventional unification\footnote{ A
heterotic string cut-off \cite{hetc}  was used in
\cite{Dudas}.}. This effective field theory study shows that that such 
scenario might be easily realized. However the precise implementation
in a string theory model needs to take care of the shortcomings mentioned 
above.

\subsection{Far and close unification}

Here we would like to discuss the possibility that logarithmic
 threshold corrections lead to unification   scale $M_X$ located much
 above (or below) the string scale. Such a scenario was  mentioned in
 \cite{caceres} and studied for the case of heterotic strings by \cite{Koun}. 
An explicit realization in open string  models appeared in \cite{bachas2}.
  
\vskip 0.5cm $\bullet$ {\em  Mechanism:}

Threshold corrections might have a logarithmic form: $\Delta_a\sim
b'_a \ln {({M_{s}R})}$ where $ b'_a$ is a numerical coefficient and
$R$ is the size  associated with a large internal dimension. In
equation (\ref{rge}) such a contribution can be seen  as a
modification of the slope of ``running'' due to the presence of matter
and leads to an apparent  unification  scale   \be M_{GUT} \sim M_s
\left( M_s R \right)^{b'_a/b_a^{(N)}}
\label{fake}
\ee where $b_a^{(N)}$ is the beta-coefficient in (\ref{rge})
connecting the last intermediary scale  and $M_s$. We see that
depending on the  sign of ${b'_a/b_a^{(N)}}$ the unification scale
might be  (further) above  or (closer) under  the string scale. For
instance  it has been proposed in \cite{bachas2} for Type I string
theories that the unification scale is of the order of Kaluza-Klein
states and which   might be very heavy leading to  $M_{GUT} \sim 1/R
\gg M_s$. In this picture one has one intermediary scale around $M_s$
where the couplings run with an $N=2$ beta-coefficients as in
\cite{BFY}.

\vskip 0.5cm $\bullet$ {\em Size of couplings:}

The discussion of the size of couplings at the string  (or physical
unification) scale is very much the same as in the accelerated
unification as both scenarios  rest on large threshold corrections.

\vskip 0.5cm $\bullet$ {\em  Advantages:}

In this picture one may perform the computation using low energy
effective action below the string scale  as if there is gauge coupling
constant unification at a higher scale $M_{GUT} \gg M_s$. This
apparent unification might have a physical origin as a cut-off due to
N=2 sector Kaluza-Klein states \cite{bachas2}.

\vskip 0.5cm $\bullet$ {\em  Shortcomings:}

The thresholds are computable only in a full $M$  or string theory
framework as  they are sensitive to the ultraviolet cut-off.

\vskip 0.5cm $\bullet$ {\em  Example: }

The scenario proposed by \cite{bachas2}.

\subsection{Hidden Unification}

It has been discovered recently that non-perturbative gauge symmetries
may arise in string  compactifications \cite{Sinst}. The associated
couplings are functions of independent moduli fields.  Some
implications for supersymmetry breaking have been discussed in
\cite{Fth}. For  the unification issue  the standard model couplings
may have arbitrary values at  the string scale. The theory (contrary
to what was used to in heterotic compactifications) makes  no
prediction of simple form of unification (if not the framework).  From
the point of view of the standard model phenomenology this seems quite
deceiving as understanding the twenty or so low energy parameters
becomes more obscure.

 A crucial  difference with ``traditional'' quantum field theories is
that in M-theory the couplings are generally vacuum expectations
values (vevs)  of (moduli) fields. Some moduli that may govern
couplings and masses of dark matter (and hidden sectors dynamics) may
be decoupled from the observable matter.  The large scale dynamics of
the universe is then governed by the variation in time  and space of
such moduli.

As we discussed before, due to large thresholds some of the gauge
couplings may evolve to very small values at the string scale
resulting in global symmetries.  In M-vacua like Type I strings, the
Newton constant also gets renormalized \cite{ABFPT}. If the threshold
corrections\footnote{Their sign depends on the number of
hypermultiplets and vector multiplets in the $N=2$ sector and may be
positive or negative.} are big then they might also drive t he
strength of gravitational interactions to very small values. M-theory
at the scale $M_s$ could become topological!.

\section{Planck, String and Compactification Scales}

We would like to discuss the inter-connections between the
four-dimensional Planck scale  $M_{Pl} \sim 1.2\times 10^{19}$ GeV, of
the string scale $M_s$ and of the volume of the internal space are
related to each other and to the ``unified'' gauge coupling at the
string scale. We will focus on two examples: M-theory on $S^1/Z_2$ and
Type I string compactifications.

\subsection{M-theory on $S^1/Z_2$}

Among the simplest four-dimensional N=1 supersymmetric vacua of
M-theory   are compactifications on  $S^1/Z_2 \times CY$ \cite
{HW,witten}, where  $S^1/Z_2$ is a segment of size $\pi \rho$ and $CY$
is a Calabi-Yau of volume  $V$. Gauge fields and matter live on the
three-branes located at each end  of the segment, while gravitons and
moduli fields ``propagate'' in the bulk.

Following \cite{witten} one may solve the equations of motion for such
 configuration as a perturbative expansion  in the dimensionless
 parameter ${\rho  M_{11}^{-3}/V^{2/3}}$. A higher orders in this
 expansion,  the  factorization in a product $S^1/Z_2 \times CY$   is
 lost. The volume of the Calabi-Yau space  becomes  a function of  the
 coordinate parametrizing the $S^1/Z_2$ segment. More precisely, the
 volumes of $CY$ seen by the observable sector\footnote{ We will use
 the subscripts $o$ for parameters of the observable sector and $h$
 for those of the hidden sector.} $V_{o}$ and the one on the hidden
 wall $V_{h}$  are given by:   \be   V_{o} = V \left(1 + \left(\frac
 {\pi}{2}\right)^{4/3} a_o {\rho  M_{11}^{-3}\over V^{2/3}}\right)
\label{volo}
\ee  and  \be  V_{h} = V \left(1 + \left( \frac {\pi}{2}\right)^{4/3}
a_h {\rho M_{11}^{-3}\over V^{2/3}} \right)
\label{volh}
\ee

where now $V$ is the (constant) lowest order value for the volume of
the  Calabi-Yau manifold and $a_{o,h}$ are model-dependent constants
\cite{nse}. Roughly speaking $a_{o,h}$ count the proportion of
instantons and five-branes on each wall.

These formulae were studied
for the standard embedding case in
\cite{witten,BD,anton,nano1,nilles}\footnote {See also \cite{wald} for
detailed discussion of the derivation of these formulae} and for the
non-standard embedding in \cite{nse,POK}. In this  last case by
putting more than half of the instantons on the hidden wall, $a_o$
becomes negative.

For a given value of $M_{11}$ we would like to determine the
corresponding values of  $V_o$, $V_h$ and $\rho$ to fit the observed
values of a unified gauge coupling $\alpha_o$ and  the Newton
constant. In the absence of a precise model, the value of the former
is unknown. We will assume that threshold corrections are small enough
so that we can take for an approximative value, the one of
$SU(3)_c$. The relevant relations are:

\be V_{o}^{-1/6}= (4\pi)^{-1/9}(\alpha_{o} f_o )^{1/6}  M_{11}
\label{bli}
\ee

and  \be  {1\over \rho} =  {16 \pi^2}{M_{11}^9 G_N \langle V\rangle}
\label{bli2}
\ee

Here $\langle V\rangle$ is the average volume of the Calabi-Yau space
on the eleven dimensional segment. The constant $f_o$ ($f_h$) is a
ratio of  normalization of the traces of adjoint representation of
$G_o$ ($G_h$)  compare to $E_8$ case \cite{witten85,nano1, nse}. There
are three different classes of solutions to consider:

\vskip 0.5cm $\bullet$ {\em Case $a_o > 0 \rightarrow$  $M_{11} \sim
10^{16}$ GeV}

Compactifications with standard embedding of the gauge connection fall
in this category (see \cite{witten}).  In these models there is an
upper limit on the size of  the $S^1/Z_2$ segment above which the
hidden sector gauge coupling blows up. If the observable sector
coupling constant is of the order of unity the corresponding lower
bound on the string scale is $M_{11}$  of the order of $10^{16}$ GeV.

This bound might be escaped  if there are  large threshold corrections
that push the unification coupling constant to much smaller values as
discussed in section 2.3.

\vskip 0.5cm $\bullet$ {\em Case $a_o = a_h =0 \rightarrow $  $M_{11}
\simgt 10^{7}$ GeV  }

In this case the only  upper limit on $\rho$ is from experiments on
modification of the Newtonian force at distances of  $\rho \simgt $ mm
\cite{caceres,IDD}. Using  $\langle V\rangle = V_o$ and $\alpha_{o}
\sim 1/10$ one obtained a lower  bound on  limit $M_{11}$ of the order
of $4\times 10^{7}$ GeV.

Some examples of characteristic size of the radii for different values
of $M_{11}$ are given in table 1.

\begin{center}
\vbox {\tabskip=0pt \offinterlineskip\def\tablerule{\noalign{\hrule}}
\def\tv{\vrule height 20pt depth 5pt}\halign to 16cm {\tabskip=0pt
plus 20mm \tv\hfill\quad#\qquad\hfill &\tv\hfill\quad#\qquad\hfill
&\tv\hfill\quad#\quad\hfill &\tv#\tabskip=0pt\cr\tablerule  $M_{11}$
in GeV &  $V_o^{-1/6}$ in GeV & $\frac {1}{\rho}$ in GeV
   
&\cr\tablerule  $2\times 10^{16}$&$1.2 \times 10^{16}$&$2 \times
10^{14}$ &\tabskip=0pt\cr $ 10^{14}$&$5.8 \times 10^{13}$&$1.5 \times
10^{8}$ &\tabskip=0pt\cr $4.2\times 10^{12}$&$2 \times
10^{12}$&$10^{3}$ &\tabskip=0pt\cr $2\times 10^{12}$&$8.6 \times
10^{11}$&$10^{2}$ &\tabskip=0pt\cr $2 \times 10^{11}$&$1.1\times
10^{11}$&$0.1$ &\tabskip=0pt\cr $4 \times 10^{10}$&$1.6\times
10^{10}$&$10^{-3}$ &\tabskip=0pt\cr $4\times 10^{8}$&$1.6\times
10^{8}$&$10^{-9}$ &\tabskip=0pt\cr $4\times
10^{7}$&$10^{7}$&$10^{-12}$ &\tabskip=0pt\cr\tablerule}}
\end{center}

\vskip 0.5truecm
\begin{center}
{\bf Table 1.}  Examples of values of approximative sizes of the
internal space radii in compactifications of $M$-theory with
$a_o=a_h=0$. We used $\alpha_o \sim \alpha_3(M_s)$ and $f_o=6$.
\end{center}

\vskip 0.5cm $\bullet$ {\em Case $a_o <0 \rightarrow$  $M_{11} \simgt$
TeV with $\rho^{-1} \ll $ TeV}

The possibility of  $a_o <0$ has been shown\footnote{For instance an
explicit three-generation $E_6$  model was exhibited in \cite{nse} and
was found to correspond to $a_o=-8$. We  take this value as typical
order of magnitude in our numerical results.} to arise in the
non-standard embedding in \cite{nse} (see also \cite{POK}).   In this
scenario, as $\rho$ increases the volume of the internal space on the
observable wall  is fixed as to fit the desired value of $\alpha_{o}$
while  the volume on  the other end of the segment increases leading
to smaller values of the corresponding coupling constant.  Typically,
$\langle V\rangle \sim \frac {V_h}{2} \gg V_o$ for  large  values of
the radius $\rho$.  Given a value of $M_{11}$ both $V_o$ and $\rho
\langle V\rangle$ can be tuned to fit the value of $\alpha_{o}$ and
$M_{Pl}$. The value of $\rho$ is then extracted from (\ref{volo}).

In table 2 we illustrate the expected sizes of the volume on the
hidden wall and the radius of the fifth dimension on some examples.

\begin{center}
\vbox {\tabskip=0pt \offinterlineskip\def\tablerule{\noalign{\hrule}}
\def\tv{\vrule height 20pt depth 5pt}\halign to 16cm {\tabskip=0pt
plus 20mm \tv\hfill\quad#\qquad\hfill &\tv\hfill\quad#\qquad\hfill
&\tv\hfill\quad#\quad\hfill &\tv#\tabskip=0pt\cr\tablerule  $M_{11}$
in GeV &  $\langle V \rangle^{-1/6}$ in GeV & $\frac {1}{\rho}$ in GeV
   
&\cr\tablerule   $ 10^{13}$&$7.7\times 10^{11}$&$5\times 10^{9} $
&\tabskip=0pt\cr $ 10^{12}$&$4.8\times 10^{10}$&$8\times 10^{7} $
&\tabskip=0pt\cr $ 10^{11}$&$3\times 10^{9}$&$1.2\times 10^{6} $
&\tabskip=0pt\cr $10^{10}$&$1.6\times 10^{8}$&$ 6\times 10^{4}$
&\tabskip=0pt\cr $5\times 10^{6}$&$2\times 10^{4}$&$2\times 10^{-2}$
&\tabskip=0pt\cr $10^{4}$&$12$&$3\times 10^{-7}$ &\tabskip=0pt\cr
$2\times 10^{3}$&$1.7$&$ 2\times 10^{-8}$ &\tabskip=0pt\cr\tablerule}}

\vskip 0.5truecm {\bf Table 2.}  Examples of values of approximative
sizes of the internal space radii in compactifications of $M$-theory
with  $a_o=- a_h=-8$. We used $\alpha_o \sim \alpha_3(M_s)$ and
$f_0=6$.
\end{center}

Larger values for $\rho$ can be obtained the following way: One starts
with a symmetric embedding i.e. putting the same number of instantons
(five-branes) on both boundaries. Then one moves by very short
distances five-branes from the observable wall. To get $\rho \sim$ mm
one needs to move one five-brane by around an Angstrom away from our
wall \footnote{ One may also see this as a fine-tuning of $a_o$ as this 
will take avalue $({x^{11}}^2 / {\pi \rho})^2$ where $x^{11}$ is the position of 
the five-brane (see \cite{LOW}).}.

In this case of non-standard embedding, as first discussed in
\cite{nse}, the  hidden observer living on the other wall could see
the new dimensions at energies ( e.g. GeV) much before the observers
on our wall (TeV). This possibility suppose however a better precision
of measurements as the interactions are weaker on his side.

Also as mentioned in \cite{nse}, at energies of the order of $GeV$ the
states in the bulk are not anymore the regular  Kaluza-Klein
states. Instead, one expects heavier modes localized on our side of
the universe which decay to lighter massive modes  localized near the
other  wall before the latter decay to hidden matter.

\subsection{Type I strings}

A  simple framework suitable to discuss both gauge and gravitational
couplings size is orbifold compactification \cite{chiralI}. In this
case the compact space  is a product of three tori $T_1 \times T_2
\times T_3$ divided by a discrete  symmetry leading to internal
volumes parametrized as $(2\pi)^2 R^2_1$, $(2\pi)^2 R^2_2$ and
$(2\pi)^2 R^2_3$  respectively.

The four-dimensional Planck mass $M_P$ and the Newton's constant $G_N$
are given by
\begin{equation}
G_N^{-1}=M_P^2 =  {{8 M_s^8 R^2_1 R^2_2 R^2_3} \over  g_s^2}
\label{mpI}
\end{equation}

and the gauge couplings of the states on the nine-branes (99) and on
the five-branes (55) are given by:

\begin{equation}
g_9^{-2} = {{M_s^6 R^2_1 R^2_2 R^2_3} \over {2 \pi g_s}} \, , \quad
\quad g_5^{-2} = {{M_s^2 R^2_i} \over {2 \pi g_s}}
\label{couplingI}
\end{equation}

where $g_s$ is the string coupling and the indices $i$ indicate around
which internal torus two of the world-volume directions are wrapped.

In case some volume $v_i$ is much smaller than the string scale, one
performs a T-duality transformations on $T_i$ which exchanges the role
of Newman and Dirichlet  boundary conditions (thus Kaluza-Klein and
winding modes). This leads to:

\begin{eqnarray}
         g_s &\rightarrow&  {g_s \over {R^2_i M_s^2}}  \nonumber \\
         R^2_i &\rightarrow & {1 \over {R^2_i M_s^4}}
\end{eqnarray}

The string scale $M_s$ is then given by: \be M_s = \left( \frac
{\sqrt{2}}{ \alpha_{o} M_P} \right)^{1/2} (R^2_1 R^2_2 R^2_3)^{-1/4}
\label{mmI}
\ee

For\footnote{ In case of large thresholds, the tree level relations
need to be modified.} $\alpha_{o} \sim$ 1, we see that to get a small
value of $M_s$ we need a large volume. One large dimension corresponds
to a $Z_2$  orbifolds while $Z_7$ requires all six dimensions to be of
the same size.

Some examples of values for the size of the radii are given in Table
3. As pointed out by \cite{AADD} the tree level relations do not allow
a single large dimension while the  string scale is lowered to $M_s
\sim$ TeV.

\vfill\eject \vbox {\tabskip=0pt
\offinterlineskip\def\tablerule{\noalign{\hrule}} \def\tv{\vrule
height 20pt depth 5pt}\halign to 17cm {\tabskip=0pt plus 20mm
\tv\hfill\quad#\qquad\hfill &\tv\hfill\quad#\qquad\hfill
&\tv\hfill\quad#\qquad\hfill &\tv\hfill\quad# \qquad\hfill
&\tv#\tabskip=0pt\cr\tablerule  $M_s$ in GeV &   $1/R_{(1)} $ in GeV &
$1/R_{(2)}$ in GeV & $1/R_{(4)}$ in GeV &\cr\tablerule   $10^{16}$&$8
\times 10^{12}$&$3 \times 10^{14}$&$2 \times 10^{15}$ &\tabskip=0pt\cr
$10^{14}$&$5\times 10^{6}$&$2\times 10^{10}$&$10^{12}$
&\tabskip=0pt\cr $10^{13}$&$5\times 10^{4}$&$2\times 10^{8}$&$5\times
10^{10}$ &\tabskip=0pt\cr $10^{12}$&$3$&$10^6$&$10^{9}$
&\tabskip=0pt\cr $10^{11}$&$3\times 10^{-3}$&$2\times 10^{4}$&$4\times
10^{7}$ &\tabskip=0pt\cr $10^{10}$&$10^{-6}$&$200$&$10^{6}$
&\tabskip=0pt\cr $10^{7}$&$2\times 10^{-15}$&$10^{-4}$&$40$
&\tabskip=0pt\cr $10^{3}$&$10^{-27}$&$10^{-12}$&$3\times 10^{-5}$
&\tabskip=0pt\cr\tablerule}}

\vskip 1.0truecm \centerline{\bf Table 3} \vskip 0.5truecm Examples of
values of parameters of the the compactification of Type I theory with
low string scale $M_s$. The cases $1/R_{(1)}$, $1/R_{(2)}$ and
$1/R_{(4)}$ correspond to anisotropic Calabi-Yau with one, two or four
dimensions with large radii. We use $\alpha_o \sim \alpha_3 (M_s)$.

\section{ Mechanisms for supersymmetry breaking:}

From a phenomenological point of view, low energy supersymmetry is
motivated by the necessity to stabilize the hierarchy  of scales
present in most of the extensions of  the standard model. If the same
motivation is invoked to set  the string scale to be as low as the
$TeV$, then it is natural to ask that  no  supersymmetry is present
and there is no need  to discuss its breaking. However, one may
insists  on supersymmetry for other reasons or consider the string
scale to lie at much higher energies: $M_s \gg$ TeV. The absence of
observation of any supersymmetric partners of standard model
particles, it is natural to demand that supersymmetry is spontaneously
broken  at energies at least of the order of the electroweak scale. In
these section we will investigate the fate of popular mechanisms for
to achieve this breaking when applied to LQGS models. In abscence of explicit
models, our discussion is deliberately made sketchy and remain at a qualitative level. Our main interest is to point out different scenarios and the challenges behind their implementation in realistic models. The latter goes beyond the scoop of this paper
.

\subsection{Gravity Mediated Supersymmetry Breaking:}

In this scenario supersymmetry breaking originates in a hidden sector
that communicates with the observable sector only through
gravitational interactions.

If all the internal dimensions are smaller than the TeV$^{-1}$ scale,
then the effective theory at  the electroweak scale is
four-dimensional.   The  supersymmetry breaking soft terms are given
by:

\be m_{soft}^2 \sim \frac {F^2}{M_P^2}
\label{soft}
\ee

where ${F^2}$ is the density of energy responsible for supersymmetry
breaking. For instance in the case  of gaugino condensation \cite{gc}
$F \sim \Lambda^3/M_P$ where $\Lambda^3$ is the vacuum expectation
value of the gaugino condensate.

To get soft-terms of the order of TeV the $F$-term has to be of the
order of:

\be \sqrt{F} \sim 10^{11} {\rm GeV}
\label{Flimit}
\ee
 
which implies $M_s \simgt 10^{11}$ GeV. This bound becomes $M_s \simgt
10^{13} {\rm GeV}$ in the case of gaugino condensation.

If the source of supersymmetry breaking is located on a hidden wall
 located at the other end of a segment with large size $R$ separating
 it from our world, the same relation (\ref{soft}) remains true. The
 large distance between  the two walls constitutes a low infrared
 cut-off that suppress the contributions of  from heavy excitations of
 bulk fields.

If $n$ internal space dimensions have  sizes  below the electroweak
scale the situation becomes more difficult. In this case the number of
states that contribute increases with energy as $(ER)^n$ leading to:

\be m_{soft}^2 \sim   \left[ 1+ \beta (ER)^n\right]  \frac {F^2}
{M_P^2}
\label{soft2}
\ee

This formula is a simple estimate of orders of magnitude. The factor
$\beta$ for instance reflects  the fact that in the bulk there are, in
addition to massive excitations of gravitons, excitations of
graviphotons and other moduli fields whose massless  partners have
been projected out in the process of supersymmetry  reduction. These
states might contribute with different strength, through both
attractive and repulsive interactions \cite{scherk, AADD}. A
difficulty in  applying this formula is to decide at which scale $E$
must be taken.

To compute the value of the soft-masses at the electroweak scale  one
may take $E$  to be $\sim$ TeV. For example the case of  $n=$1 or 2
dimensions of size  $10^{-3}$ eV the limits on  on $\sqrt{F}$ become
of the order of $3 \times 10^6$ GeV and   TeV.

If instead $E$ has to run to the infrared cut-off, then $E \sim 1/R$
and one recovers the result of the four-dimensional case.

If supersymmetry breaking  originates  in $F$-terms for moduli
fields. In general these moduli have non-universal coupling to matter
which might lead to non-universal soft terms on the observable sector.

\subsection{Gauge Mediated Supersymmetry Breaking:}

This scenario \cite{gm} assumes that supersymmetry is broken in a
secluded sector of the theory.  Some states  are considered to be
charged under both the observable and  secluded sectors and thus
mediate the supersymmetry breaking through gauge interactions.

Within our picture of walls (three-branes) separated by the  bulk, we
may consider the following three cases:

\vskip 0.5cm $\bullet$ {\em Secluded and observable sectors on the
same wall:}

In type I strings, this might for example if on the same point one
sector arises from nine-branes  (99) while the standard model lives on
fivebranes (55)  (or  seven-branes and three-branes after
$T$-duality).  The sector communicating the supersymmetry breaking
would then be the (59) (or  (73) after $T$-duality)  open strings that
have one end on the five-branes and another on the nine-branes.

In this case the states in the  bulk do not participate to the
supersymmetry breaking mediation. The computation and results are very
much standard and lead to a mass for gauginos of the order of: 
\be
m_{{1/2_a}}\sim k_a \frac  {\alpha_a} {4\pi} N  M_{ms}
\label{gmsb1}
\ee

and for scalar masses of the order of:

\be m_{o_i}^2\sim  \Sigma_{a=1}^3 c_a  k_a^2\frac {\alpha_a^2} {(4\pi)^2} \left[{\lambda_a {N}+\gamma_a N^2}\right]
M_{ms}^2
\label{gmsb2}
\ee

where $M_{ms}$ and $N$ are the mass scale and  the number of
messengers. the coefficients $c_a, \lambda_a$ and $\gamma$ are model dependent
.  For simplicity, we have assumed that their mass splitting is of the
order of $M_{ms}$.  The latter must satisfy  $10\, \, {\rm TeV} \simlt N\,
M_{ms}  \simlt M_s$ which implies (for low values of $N$) a string
scale $M_s \simgt $ 10 TeV.

For a string scale $M_s$ of the order of TeV,  a large $N$ is
necessary to not get too small soft terms. This usually enhances the
difference of masses between  gauginos and scalars. One might speculate
that 
Kaluza-Klein states who became massive due gauge symmetry breaking
using Wilson lines would play the role of messengers. hower, outside the Scherk-Schwarz mechanism it is not clear how to generate mass splitting for these states.

\vskip 0.5cm $\bullet$ {\em Secluded sector in the bulk and observable
sectors on the  wall:}

In type I strings, this might arise if the dimension with large size
is one of the directions  orthogonal to the five-brane where the
observable sector resides. The secluded sector arises from
nine-branes while the  messengers are (59) open strings.

If the distance between the walls is smaller than $\sim M_{ms}$  then
the result is identical to the one in the previous section. However,  
if the size $R$
of the separation becomes bigger, then the four-dimensional coupling
becomes  very small and it is difficult to resort to gauge dynamics to
generate supersymmetry breaking  of order of $M_{ms}$. An alternative would be
that 
supersymmetry is broken by a Sherck-Schwarz mechanism.  We
discuss this issue in the next section.

\vskip 0.5cm $\bullet$ {\em Secluded  and observable sectors on two
opposite  walls:}

Finally  supersymmetry might be broken on the opposite wall and later
mediated through additional gauge interactions present in the bulk
under which quarks  and leptons are charged. This possibility has been
studied in \cite{Peskin} in five-dimensions.

The messengers scale $M_{ms}$ plays the role of a cut-off in the loops
responsible of the mediation  of supersymmetry breaking. Thus for a
distance between the walls $R < M_{ms}^{-1}$ Kaluza-Klein states are
not excited and the result is the same as if the space was
four-dimensional.

When the radius of the fifth dimension encreases $R > M_{ms}^{-1}$
Kaluza-Klein excitations of the gauge bosons are excited. Thus the
gauge couplings get contributions from $(R  M_{ms})^n$ states leading
to the changement: \be \alpha_a \rightarrow   \frac {\alpha_a}{ (R
M_{ms})^n}
\label{alp}
\ee

This simple scenario is  not appealing for large $R$ from the
phenomenological point of view. The gaugino masses on the observable
wall have to be generated at higher orders, and even the scalar masses
are small because the four-dimensional gauge coupling in the bulk
should be suppressed by the large volume. 

Finally, the case of the standard model residing in the bulk is very similar 
to the  case of orbifold compactifications of heterotic strings. One is faced in this case with the problem of power law running of standard model couplings.

\subsection{Scherk-Schwarz mechanism:}

This mechanism requires the existence of a symmetry group $G_{ss}$
that does not commute with supersymmetry.  The members of the same
supersymmetric multiplet have different charges $q_i$ under
$G_{ss}$. Instead of the  usual periodic conditions when going around
some  direction of the internal  space  of   circle of radius $R$ ,
some states transform non-trivially under $G_{ss}$. In the simplest
case, the result for states with mass:
 
\be m_n^2= \frac {n^2}{R^2} +  l^2 {R^2} M_s^4
\label{kk}
\ee

is to shift  $n\rightarrow n+q_i$  or $l\rightarrow l+q_i$.  This
creates a splitting inside each multiplet and thus it breaks
supersymmetry.  The simplest example for $G_{ss}$ would be  R-parity
($q=o$ for standard particles and $q=1/2$ for sparticles: gauginos,
sleptons, squarks and Higgsinos). Another commonly used symmetry is
the spin: $q=s$ which is integer  for bosons and half-integer for
fermions. In this case fermionic matter, leptons and quarks, have to
be identified with ``twisted states'' living on branes (or
orientifolds) orthogonal to the $z$ direction.

Within our picture of a world made of  walls and bulk, the
implementation of the Scherk-Schwarz mechanism leads to many different
scenarios:

\vskip 0.5cm $\bullet$ {\em Gravity mediated Scherk-Schwarz
supersymmetry breaking}:

The first possibility is to consider shift in momenta or winding in a
direction orthogonal to world--volume of the brane on which the
standard model states live.  At tree level only the states propagating
in the bulk feel supersymmetry  breaking: a mass splitting between
supersymmetric partners is generated in the hidden sector.

If there are no large dimensions lying under the TeV scale then
supersymmetry breaking  is communicated to the observable sector
through four-dimensional gravitational interactions.  The resulting
soft-terms are of the order of \cite{AMQ} $m_{soft} \sim \frac {1}{R^2
M_{Pl}}$ or  $\frac {R^2 M_s^4}{ M_P}$ depending if  the shift was made on
the momenta or windings. In the case of presence of $n$ internal
dimensions with larger radii $r$ i.e. $r^{-1} \simlt TeV$ then the
strength of gravitational  strength changes with energy, leading to a
multiplicative factor of order of ${(Er)^n}$.

\vskip 0.5cm $\bullet$ {\em Gauge mediated Scherk-Schwarz
supersymmetry breaking:}

A different scenario may be illustrated on the following example:
Suppose that the standard model lives on five-branes and a hidden
sector arises from the ninebranes. There are (59) strings with one end
on the five-branes and one on the nine-branes. The corresponding states
 are charged under both groups.

If the non-trivial periodic condition is on a  direction orthogonal to
the fivebrane.  Only the ninebranes will feel the supersymmetry
breaking at tree level. However this might be communicated to the
five-brane. First, the (59) open strings will have splittings due to
radiative corrections from (99) sector gauge symmetry. Then  the
(59) open strings will generate soft breaking in the observable
sector. The scale $1/R$ might lie much higher than the TeV if the
gauge coupling in the (99) brane is small.

\vskip 0.5cm $\bullet$ {\em Direct  Scherk-Schwarz breaking:}

Another  possibility that has been studied in \cite{Ant} is that the
coordinates affected by the Scherk-Schwarz mechanism are parallel to
the world-volume of the brane on which the  standard particles
reside. In this case soft masses generated for  the standard particles
are of the order  of $1/R$.

Existing N=1 Type I string compactifications seems to lead only to
singlets  from twisted states. To avoid giving large masses to
standard model fermions and bosons the charge $q$  could be
associated to the $R$-parity charge.  The boundary conditions are thus
universal: \be m_0= m_{1/2}=m_{3/2}\sim 1/R \sim TeV
\label{blue}
\ee where $m_0$, $ m_{1/2}$ and $m_{3/2}$ denote the scalar, gaugino
and gravitino soft-masses.

Other possibilities  could be engineered for this kind of
compactifications if only a part of the standard model lives on one
stack of five-branes orthogonal to  the affected direction. For example:

- $SU(3)$ gauge  symmetry might arise on ninebranes parallel
to the affected direction. In contrast $SU(2)\times U(1)$ would arise
 from fivebranes orthogonal to it. Leptons and Higgs would live in 
this fivebrane sector and
quarks doublets would originate from open strings stretching between the fivebranes and ninebranes. Only the  gluinos would have a tree level soft masses.

- If  only $SU(2)$ arises from the nine-branes then only the
  corresponding gauginos have soft masses.

- If $U(1)$ arises from the nine-brane sector then only the bino has
a tree level soft mass.

- If $SU(2) \times U(1)$ arises from nine-branes then the squarks and
 gluinos have vanishing soft masses.

Finally there is the possibility where the theory arises from an
orbifold compactifications (of M-theory) with the matter fields in the
twisted sector while the gauge bosons are in the bulk (untwisted
sector). The Higgs fields might be chosen in the twisted or untwisted
sectors \cite{AMQ}.  The first case leads to $m_0=m_3=0$ and
$m_{1/2}=m_{3/2}\sim 1/R \sim$ TeV.  while the later gives $m_o=0$ and
$m_3=m_{1/2}=m_{3/2}\sim 1/R \sim$ TeV.

All these scenarios assume a large dimension $1/R \sim TeV$.  The
supersymmetry breaking is communicated from one set of the fields to
the other one through gauge interactions.

The computation  of the sparticle spectrum assumes that one can evolve
the the coupling  constants above the TeV scale. In which case for low
$M_s$ the running introduces small non-universalities.  This is
possible if the KK states associated  with the large radius do not
contribute to the running. This was argued to be the case if the  KK
states are in (spontaneously broken) $N=4$ multiplets \cite{Ant}. If they are
instead in $N=2$  representations they  generically lead to large
(power law increasing) corrections and can not be computed reliably in
a  field theory framework.

Low energy consequences of these scenarios will be
discussed elsewhere \cite{elena}.

\section{ Other phenomenological implications and the preferred value for the string scale:}

We have discussed above scenarios some implications of the LQGS models
for the unification and supersymmetry breaking scenarios. Here we
would like to comment on other  possible phenomenological implications.

$\bullet$ {\em Dark matter:}

A hidden wall is a candidate to contain an important fraction of dark
 matter.  In the M-theory context this possibility has
 appeared\footnote{ The phenomenology is  similar to the shadow matter
 that has been studied for instance in \cite{KT}.}  (to our knowledge)
 for the first time in \cite{BEN} and discussed in some details in
 \cite{nse}. On a simple example of inflation, it was shown in
 \cite{kar1} how the dynamics on the two walls, observable and hidden,
 can be interconnected.  

Here we wish to discuss the brane scenario for another issue:
that the cosmological constant is simulated by some field with variable mass
or ``quintessence''\cite{quinte}.

Suppose that we try to fit the expansion rate of
 the universe can  by including dark matter with a variable mass
 \cite{varmass}. In the perturbative heterotic string scenario, this typically leads to varying the strength of the gravitational and gauge 
coupling on the  observable world. There are strong constraints on 
such variation that make this scenario unlikely.

Let's consider compactifications of Type I on orbifolds. Suppose that
there are stacks of  fivebranes wrapped around each of the tori.  The
standard model may arise from a fivebrane with two world-volume
internal dimensions wrapped around  the internal torus $T_1$ with
constant volume $v_1$. Now, let's suppose that the volumes $v_2$ of
$T_2$ and $v_3$ of $T_3$ vary with time  such that the product $v_2
v_3$  remains constant. The gauge coupling constants on the observable
world depend on $v_1$ only while  the gravitational coupling depends
on the product of $v_1 v_2 v_3$, so both are constant in time while
dark matter couplings depend on the volume of the internal space and
thus vary with time. While probably present, such a scenario has not
been found  in known heterotic string compactifications
\cite{varmass}.  The mass of dark matter is very model dependent is
 but one expects it to depend on the
gauge coupling thus leading to dark matter with variable mass.
For instance if the hidden dark matter is made of confined hidden particles, 
then there mass is governed by the confinement scale. The latter 
is obviousely varying with the strength of the tree level coupling constant.
This phenomenon  seems to be allowed by Type I string theory.

In the context of Ho\v rava-Witten type of models, dark matter with
variable mass might be obtained by taking one or a set of five-brane
and arranging that they move in the  fifth dimension separating the
two boundaries.  A judicious choice of five--branes  allows the
coupling constant on the observable wall to remain constant.
For instance one could take a couple of five-branes: one at $\pi \rho \cos{z(t)}$ and the other $\pi \rho \sin{z(t)}$ where $z(t)$ is a slowly varying phase.

\vskip 0.5cm $\bullet$ {\em Neutrino masses:}

Recent data from different experiments suggest existence of
oscillations of between  different neutrinos. Such processes require
that the neutrinos are massive  In a minimal scenario, one tries to
build a mass matrix with three neutrinos  which allows to fit the data
from solar and atmospheric neutrinos experiments.

Let us first discuss this issue in the left-right class of models
presented in the appendix. The neutrino masses are given by :

\be m_{\nu_i} \sim \frac {m_{Di}^2}{M_R}
\label{seesaw}
\ee where $m_{Di}$ are Dirac neutrino masses and it is a free
parameter. For $M_R \sim 10^8$ GeV, a neutrino mass of $\sim$ eV
corresponds to $m_{Di} \sim 0.1$ to 1 GeV.

Another possibility\footnote{ Relating the neutrino mass to existence
of  extra-dimensions at scales of the order of $10^{12}$ GeV is under
investigation with J. Ellis \cite{ellis}.} is to rely on the violation
of global symmetries by quantum gravitational effects \cite{lept}. 
This is arises in string theory du to the presence of heavy (oscillators) modes
with interactions that violate these global symmetries.
For
instance the violation of lepton number would  lead to operators of
the form $\frac {1}{M_s}LLHH$.  If $M_s$ is in the region of $10^{11}-10^{13}$
GeV then the neutrino masses might be naturally of the order of eV (depending on the precise value of the coefficient of this operator).

Finally, it was proposed that a modulino might play the role of a
sterile neutrino \cite{Smi}. The modulino-neutrino mixing would arise
from from $R$-parity bilinear terms of the form $\mu LH$ through the
dependence of $\mu$ on the modulus $S$. To get light neutrinos one
takes $\mu \sim 1$ to 10 GeV. This values imply that the
modulino-neutrino mixing mass will be of the order of $eV$ for
$\langle S \rangle$ of the order of $\langle S \rangle \sim M_s \sim
10^{11}$ to  $10^{12}$ GeV. For scenarios where the modulino is light enough
this might explain the different neutrino anomalies.

\vskip 0.5cm $\bullet$ {\em A preferred value for the string scale?:}

M-theory as known today seems to allow arbitrary values for the string
scale. Only experimental limits  seem to imply that it is not lower
than the TeV. A TeV scale is certainly exciting as it could be probed
at future colliders. However there are no experimental indications
supporting the existence of such a scale. Three other scales might be
considered as more motivated from our observations: $10^{19}$ GeV
which is the natural scale, $10^{16}$ GeV if one believes that  at
this scale all interactions should unify (as suggested by LEP) and
finally we suggest $10^{10}$--$10^{14}$ GeV centered around $10^{12}$
GeV which is our preferred value. In fact this scale appears
naturally when one tries to explain many experimental observations as
the neutrino masses discussed above  or  the scale for axion
physics. For instance the breaking of Peccei-Quinn symmetry\footnote
{ The proposal to solve the axion  problem by decreasing the string
scale was made by \cite{choi} then more recently by
\cite{ADD2}. However they both considered different values of $M_s$.}
is constraint by cosmological and astrophysical bounds to  be roughly
in the region of $10^{10}$--$1 0^{12}$ GeV. The presence of quantum
gravitational effects at this scale due to its identification with
$M_s$ may be responsible of the breaking of the symmetry. Moreover,
the observed ultrahigh energy cosmic rays might may just originate at
the string scale. One can speculate on their origin as coming from
decay of long lived massive string modes, or p-branes wrapped around
some internal space direction.

\section{Conclusions:}

In summary, in this paper we have considered many phenomenological
aspects of LQGS models and we obtained in our opinion many interesting
new results. For instance:

$\bullet$ In contrast with the claims of recent litterature,
unification in  LQGS models can be achieved in different ways. For
certain  values of the string scale $M_s$,  this can be achieved
without introduction of ad-hoc exotic matter, and in most cases one
does not need to appeal to  threshold effects as in
accelerated unification. However, if $M_s$ becomes of the order of the
TeV we argue that unification should be studied within a full string
theory framework.

$\bullet$ We have exhibited compactifications of Horava--Witten
M-vacua that lead to an eleven--dimensional scale of the order of TeV
while only one internal dimension  has a size  in the $10^{-5}$ to 1
mm region. We illustrated examples for the size of  the radii if the
internal space dimensions when the string scale varies from TeV to
Planckian energies.

$\bullet$ We have studied different scenarios for supersymmetry
breaking and  pointed out the problems when trying to apply them to
phenomenological considerations.

$\bullet$ Finally, we have addressed some phenomenological issues:
dark matter, neutrino masses,  axion scale and ultra-high cosmic
rays. While we believe a string scale at the TeV energies is
appealing  experimentally, we suggest that  the experimental data
might seem more  natural if $M_s$ is in the range of
$10^{10}$--$10^{13}$ GeV.
 
In this paper, we have began the study of some implications of having
a low scale for quantum  gravitational effects. In the absence of
concrete models, many of the issues  were discussed at a  qualitative
level. We believe that many of them merit to be studied further.

{\bf Note added} When this manuscript was in preparation
ref. \cite{tye2} appeared that overlaps with part of Sections 4.1 and
4.3.

\section{Acknowledgements:} I  wish  to thank I. Antoniadis for comments on an earlier version of the manuscript and 
C. Bachas for e-mail communications related to his work.  I also would
 like to thank E. Accomando, S. Davidson, M. Duff, E. Dudas, J. Ellis,
 E. Kiritsis, C. Kounnas, M. Porrati and  S. Stieberger for useful
 discussions. This work is supported by a John Bell scholarship from
 the World Laboratory.

%%%%%%%%%%%%%%%%%%%%%%%%%%%%%%%%%%%%%%%%%%%%%%%%%%%%%%%%%%%%%%%%%%%%%%%%%%%%%%%


\begin{thebibliography}{15}

%%%%%%%%%%%%%%%%%%%%%%%%%%%%%%lowM_s%%%%%%%%%%%%%%%%%%%%%%%%%%%%%%%%%%%%%%%%%%%%

\bibitem{witten} E. Witten, \Journal {\NPB} {471}{135} {1996};


%%%%%%%%%%%%%%%%%%%%%%%%%%%%LEP%%%%%%%%%%%%%%%%%%%%%%%%%%%%%%
\bibitem{LEP1} For the prediction of the unification scale see:
S. Dimopoulos, S. Raby, and F. Wilczek,\Journal{\PRD}
{24}{1681}{1981}; N. Sakai, {\em Zeit. Phys.} {\bf C 11}153
(1981);L. E. Ib{\'a}{\~ n}ez and G. G. Ross, \Journal{\PLB} {105}
{439}{1981}; M. B. Einhorn and D. R. T. Jones,
\Journal{\NPB}{196}{475} {1982}; W. J. Marciano and G. Senjanovi{\'c},
\Journal{\PRD} {25}{3092} {1982}.


\bibitem{LEP2} For the analysis of LEP data see for example: \\
U. Amaldi, A. Bohm, L.S. Durkin, P. Langacker, A.K. Mann,
W.J. Marciano, A. Sirlin and H.H. Williams, \Journal{\PRD}{36}
{1385}{1987}; \\ G. Costa, J. Ellis, G.L. Fogli, D.V. Nanopoulos and
F. Zwirner, \Journal{\NPB}{297}{244}{1988};\\  J. Ellis, S. Kelley,
and D. V. Nanopoulos,  \Journal{\PLB}{249}{441}{1990};
\Journal{\PLB}{260}{131}{1991}; \Journal{\NPB}{373}{55}{1992};\\
P. Langacker and M. Luo,  \Journal{\PRD}{44}{1991}{817};\\
U. Amaldi, W. de Boer, and H. F{\"u}stenau, \Journal{\PLB}{260}
{447}{1991};\\ C. Giunti, C.W. Kim and  U.W. Lee, \Journal{\MOD} {A6}
{1745}{1991}.

%%%%%%%%%%%%%%%%%%%%%%%%%%%%%%%%bounds%%%%%%%%%%%%%%%%%%%%%%%%%%%%%%%%%%%%%%%%%%%%%%%

\bibitem{AB}  I. Antoniadis and K. Benakli, \Journal{\PLB} {326}
{1994} {69}.



\bibitem{caceres} E. Caceres, V. S. Kaplunovsky, I. Mandelberg,
 \Journal{\NPB}{493}{73}{1997}.

%%%%%%%%%%%%%%%%%%%%%%%%%%TeV%%%%%%%%%%%%%%%%%%%%%%%%%%%%%%%%%%%
\bibitem{Ant}     I. Antoniadis, \Journal {\PLB}{246}{377}{1990};


\bibitem{lykken} J. Lykken,  \Journal{\PRD} {54} {3693}{1996}.

\bibitem{bachas1} C. Bachas, (1995) unpublished.

\bibitem{ADD}  N. Arkani-Hamed, S. Dimopoulos, and G. Dvali,
      hep-ph/9803315.

%%%%%%%%%%%%%%%%%%%%%%%%%%TeV%%%%%%%%%%%%%%%%%%%%%%%%%%%%%%%%%%%
\bibitem{AADD} I. Antoniadis, N. Arkani-Hamed, S. Dimopoulos and
G. Dvali, hep-ph/9804398.

\bibitem{tye}  G. Shiu and S.-H.H. Tye,  hep-th/9805157.

%%%%%%%%%%%%%%%%%%%%%%%%%%LHC%%%%%%%%%%%%%%%%%%%%%%%%%%%%%%%%%%
\bibitem{ABQ} I. Antoniadis, K.  Benakli and M. Quir{\'o}s,  \Journal
    {\PLB} {331}{313} {1994}.



%%%%%%%%%%%%%%%%%%%%%%%%%%Unification0%%%%%%%%%%%%%%%%%%%%%%%%%%%%%%%%%%%

\bibitem{Dudas}  K. Dienes, E. Dudas and T. Gherghetta, hep-ph/9803466.

\bibitem{bachas2} C. Bachas, hep-ph/9807415.



\bibitem{Irev} J.  Polchinski, S. Chaudhuri and C.V. Johnson,
hep-th/9602052;\\ J. Polchinski,  hep-th/9611050; \\ C.Bachas,
hep-th/9806199.

%%%%%%%%%%%%%%%%%%%%%%%%%%HW1%%%%%%%%%%%%%%%%%%%%%%%%%%%%%%%%%%%

\bibitem{HW} P. Ho\v rava and E. Witten, \Journal {\NPB} {460}{506}
    {1996};  \Journal {\NPB} {475}{94} {1996}.



\bibitem{stelle} A. Lukas, B.~A. Ovrut, K.S. Stelle and D. Waldram,
     hep-th/9803235, hep-th/9806051;\\ The world as a brane was
     proposed by:\\ V.A. Rubakov and M.E. Shaposhnikov, \Journal
     {\PLB} {125}{136}{1983}.
 


%%%%%%%%%%%%%%%%%%%%%%%%%%nse%%%%%%%%%%%%%%%%%%%%%%%%%%%%%%%%%%%
\bibitem{nse} K. Benakli,  hep-th/9805181.


\bibitem{POK}  Z. Lalak, S. Pokorski and S. Thomas, hep-ph/9807503.

\bibitem{LOW} A Lukas, B.A. Ovrut and D. Waldram; hep-th/9808101. 

\bibitem{Stie} S. Stieberger, hep-th/9807124.

%%%%%%%%%%%%%%%%%%%%%%%%%%gauginocond%%%%%%%%%%%%%%%%%%%%%%%%
\bibitem{gc} H.-P.~Nilles, \Journal {\PLB} {115} {193}{1982}; and
\Journal {\NPB} {217}{366} {1983};\\ S.~Ferrara, L.~Girardello and
H.-P.~Nilles, \Journal {\PLB} {125} {457}{1983};\\ J.-P.~Derendinger,
L.E.~Ib\'a\~nez and H.-P.~Nilles, \Journal {\PLB} {155}{65} {1985};\\
M.~Dine, R.~Rohm, N.~Seiberg and E.~Witten, \Journal {\PLB} {156}
{55}{1985};\\ C.~Kounnas and M.~Porrati, \Journal {\PLB} {191}{91}
{1987}.

\bibitem{Mgc}  J. Ellis, Z. Lalak, S. Pokorski and W Pokorski,
hep-ph/9805377;\\ A. Lukas, B. A. Ovrut and  D. Waldram, \Journal
{\PRD}{57}{7529}{1998};\\ Z. Lalak and S. Thomas, \Journal
{\NPB}{515}{55}{1998};\\ T. Li, J.L. Lopez and D.V. Nanopoulos,
{\PRD}{56}{2602}{1997};\\ H.P. Nilles, M. Olechowski and M. Yamguchi,
\Journal {\PLB}{415}{24}{1997};\\ D. Bailin, G.V. Kraniotis and
A. Love, hep-ph/9803274;\\ K. Choi, H.B. Kim and C. Munoz,  \Journal
{\PRD}{7521}{1998};\\ S. Abel, C. Savoy; hep-ph/9809498.

%%%%%%%%%%%%%%%%%%%%%%%%%%gm%%%%%%%%%%%%%%%%%%%

\bibitem {gm} M. Dine and A. E. Nelson, \Journal {\PRD} {48} {1277}
{1993}; \\ an early implementation in string theory can be found in :
I. Antoniadis and K. Benakli, \Journal {\PLB}{295}{219}{1992};
Erratum-ibid. B {\bf 407} 449 (1997);  \\ For a review and further
references see: G.F. Giudice and R. Rattazzi, hep-ph/9801271.

\bibitem{Peskin}  E.A. Mirabelli and M.E. Peskin, hep-th/9712214.

%%%%%%%%%%%%%%%%%%%%%%%%%%ss%%%%%%%%%%%%%%%%%%%

\bibitem{ss}  J. Scherk and J.H. Schwarz,
              \Journal{\PLB}{82}{60}{1979};\\ E. Cremmer, J. Scherk
              and J.H. Schwarz,\Journal{\PLB}{84}{83}{1979};\\
              P. Fayet, \PLB{159}{85}{121};
              \Journal{\NPB}{263}{649}{1986}.


\bibitem{ssf}  C.~Kounnas and M.~Porrati, \Journal{\NPB} {310}
{355}{1988};\\ 
I. Antoniadis, C. Bachas, D. Lewellen and
T. Tomaras,\Journal {\PLB}  { 207} {441}{1988};\\ 
S.~Ferrara, C.~Kounnas, M.~Porrati and F.~Zwirner, \Journal{\NPB}{318}
{75}{1989};\\ 
C.~Kounnas and B.~Rostand, \Journal{\NPB} {341}{641}
{1990};\\ 
I.~Antoniadis and C.~Kounnas, \Journal{\PLB} {261} {369}{1991}.

\bibitem{ssf2}
I. Antoniadis,  Proc. PASCOS
Symposium, Boston (World Scientific, Singapore, 1991) p. 718;\\
K. Benakli,  \Journal
{\PLB}{386}{106}{1996}. 


\bibitem{ssf3}
I. Antoniadis and M. Quir\'os, \Journal {\PLB}{392}
{61} {1997};  \Journal {\PLB} {416} {327}{1998};\\ 
E. Dudas and C. Grojean, \Journal {\NPB} {507}{553}{1997};\\ 
E. Dudas, \Journal {\PLB} {416} {309}{1998};\\ 
A. Pomarol and M. Quir\'os,  hep-ph/9806263;\\ 
I. Antoniadis, E. Dudas and A. Sagnotti,  hep-th/9807011.

\bibitem{ssf4}
C. Bachas, hep-th/9503030;\\ 
H.P. Nilles and
M. Spalinski,  \Journal {\PLB}{392}{67}{1997}\\ 
I. Shah and S. Thomas,
\Journal {\PLB}{409}{198}{1997}.


%%%%%%%%%%%%%%%%%%%%%%%%%%dm%%%%%%%%%%%%%%%%%%%%%%%%%%%%%%%%%%%
\bibitem{BEN} K. Benakli, J. Ellis and D. V. Nanopoulos,
hep-ph/9803333.

\bibitem{ADD2}  N. Arkani-Hamed, S. Dimopoulos, and G. Dvali,
      hep-ph/9807344.


\bibitem{ADMR} I. Antoniadis, C. Mu\~noz and M. Quiros, {\it Nucl. Phys. } {\bf B397} (1993) 515;
 N. Arkani-Hamed, S. Dimopoulos and J. March-Russell, hep-th/9809124; R. Sundrum, hep-ph/9807348.



%%%%%%%%%%%%%%%%%%%%%%%%%%FININTRO%%%%%%%%%%%%%%%%%%%%%%%%%%%%%%%%%%%%%







%%%%%%%%%%%%%%%%%%%%%%%%%%%%%%%%%Convunif%%%%%%%%%%%%%%%%%%%%%%%%%%%%%%%%%


\bibitem{ross} For a review see G.G. Ross, {\em Grand Unified
Theories} (Benjamin/Cummings, Menlo Park, California, 1985).

\bibitem{DM} B. Dutta and R.N. Mohapatra, hep-ph/9804277.

%%%%%%%%%%%%%%%%%%%%%%%%%%ratnification%%%%%%%%%%%%%%%%%%%%%%%%%%%%%%%%%%%

\bibitem{ibanez} J.A. Casas and C. Mu\~noz, \Journal {\PLB}
          {214}{543}{1988}; L. Ib\'a\~nez, \Journal
          {\PLB}{318}{73}{1993}.

\bibitem{BS} K. Benakli and G. Senjanovi\'c, \Journal{\PRD}{54}{5734}{1996}.

\bibitem{DFM}  K.R. Dienes, A.E. Faraggi, and J. March-Russell,
      \Journal {\NPB}{467}{44}{1996}.


%%%%%%%%%%%%%%%%%%%%%%%%%%%%%%%%%typeIcutoff%%%%%%%%%%%%%%%%%%%%%%%%%%%%%%%%%%%%%
\bibitem{WL} K. Benakli,  \Journal {\PLB}{386}{106}{1996};\\

\bibitem{BF} C. Bachas and C. Fabre, \Journal {\NPB}{476}{418}{1996}.

%%%%%%%%%%%%%%%%%%%%%%%%%%%%%%%%%hetcutoff%%%%%%%%%%%%%%%%%%%%%%%%%%%%%%%%%%%%%

\bibitem{hetc} V.S. Kaplunovsky, \Journal {\NPB}{307}{145}{1988};
Erratum: {\em ibid.} B{\bf 382} (1992);  E. Kiritsis and C. Kounnas,
\Journal {\NPB}{442}{472}{1995}.

\bibitem{Koun} E. Kiritsis, C. Kounnas,
P.M. Petropoulos and  J. Rizos, \Journal {\PLB}{385}{87}{1996};
hep-th/9807067.


%%%%%%%%%%%%%%%%%%%%%%%%%%%%%%%%%hetcutoff%%%%%%%%%%%%%%%%%%%%%%%%%%%%%%%%%%%%%

\bibitem{BFY} C. Bachas, C. Fabre and T. Yanagida, \Journal
{\PLB}{370}{49}{1996}.


\bibitem{Sinst} E. Witten, \Journal{\NPB}{460}{541}{1996}; M.J. Duff,
R. Minasian and E. Witten, \Journal{\NPB}{465}{413} {1996};
P. Candelas and H. Skarke, \Journal {\PLB}{413}{63}{1997}.

\bibitem{Fth} V.S. Kaplunovsky and J. Louis, \Journal {\PLB}{417}
{45}{1998}.




%%%%%%%%%%%%%%%%%%%%%%%%%%%%%%%%%hetcutoff%%%%%%%%%%%%%%%%%%%%%%%%%%%%%%%%%%%%%


\bibitem{ABFPT} I. Antoniadis, C. Bachas, C. Fabre, H. Partouche and
T.R. Taylor, \Journal {\NPB} {489}{160}{1997}.

%%%%%%%%%%%%%%%%%%%%%%%%%%%%%%%%%hw%%%%%%%%%%%%%%%%%%%%%%%%%%%%%%%%%%%%%

\bibitem{wald} A. Lukas, B.A. Ovrut and D. Waldram,  hep-th/9710208.

\bibitem{BD} T. Banks and M. Dine,\Journal{\NPB}{479}{173}{1996}.



\bibitem{anton} I. Antoniadis and M. Quiros, \Journal{\PLB} {416}
{327}{1998}.

\bibitem{nano1} T. Li, J.L. Lopez and D.V. Nanopoulos,  \Journal{\MOD}
{A12} {2647}{1997}.


\bibitem{nilles} H.P. Nilles, M. Olechowski and M. Yamaguchi,
  hep-th/9801030.
%%%%%%%%%%%%%%%%%%%%%%%%%%f0%%%%%%%%%%%%%%%%%%%%%%%%


\bibitem{witten85} E. Witten, \Journal{\PLB}{155}{151}{1985}.

%%%%%%%%%%%%%%%%%%%%%%%mm%%%%%%%%%%%%%%%%%%%%%%%%%%%

\bibitem{IDD} I. Antoniadis, S. Dimopoulos and G. Dvali; \Journal
{\NPB}{516}{70}{1998}.


%%%%%%%%%%%%%%%%%%%%%%%Type I%%%%%%%%%%%%%%%%%%%%%%%%%%%

\bibitem{chiralI} see e.g.:\\ C. Angelantonj, M. Bianchi, G. Pradisi,
A. Sagnotti and Ya. S. Stanev, \Journal {\PLB} {385} {96}{1996};\\
Z. Kakushadze and  G.  Shiu, \Journal {\PRD} {56} {3686}  (1997);\\
G. Zwart, hep-th/9708040~;\\  G. Aldazabal, A. Font, L. E. Ibanez and
G. Violero, hep-th/9804026~;\\ Z. Kakushadze, G.  Shiu and
S.-H.H. Tye,  hep-th/9804092~;\\  J. Lykken, E. Poppitz and
S. P. Trivedi, hep-th/9806080.



\bibitem{scherk} J. Scherk, \Journal {\PLB} {88} {265}{1979}.


%%%%%%%%%%%%%%%%%%%%%%%%%%ss%%%%%%%%%%%%%%%%%%%%%%%%

\bibitem{ADS} I. Antoniadis, E. Dudas and A. Sagnotti,
hep-th/9807011.


\bibitem{AMQ} I. Antoniadis, C. Mu\~noz, and M. Quir\'os,
\Journal{\NPB}{397}{515}{1993}




%%%%%%%%%%%%%%%%%%%%%%%%%%elena%%%%%%%%%%%%%%%%%%%%%%%%

\bibitem{elena} E. Accomando and K. Benakli, in preparation.





%%%%%%%%%%%%%%%%%%%%%%%%%%dm%%%%%%%%%%%%%%%%%%%%%%%%

\bibitem{KT} see for example E. W. Kolb, D. Seckel and M. S. Turner,
\Journal{\em Nature}  {314}{1985}{415}.


\bibitem{kar1} K. Benakli,  hep-th/9804096.

\bibitem{quinte} R. Caldwell, R. Dave and P.J. Steinhardt, 
\Journal{\PRL}{80}{1582}{1998}.

\bibitem{varmass} see for e.g. : G.W. Anderson and S.M. Carroll,
astro-ph/9711288;\\ J.A. Casas, J. Garcia-Bellido and M. Quiros, {\em
Class. Quant. Grav.} {\bf 9}, 1371 (1992).


%%%%%%%%%%%%%%%%%%%%%%%%%%See-Saw %%%%%%%%%%%%%%%%%%%%%%%%
\bibitem{seesaw} M.~Gell-Mann, P.~Ramond, and R.~Slansky, \newblock in
{\em Supergravity}, edited by P.~van Niewenhuizen and D.~Freedman,
Amsterdam, 1979, North Holland; \\ T.~Yanagida, \newblock in {\em
Workshop on Unified Theory and Baryon number in the Universe}, edited
by O.~Sawada and A.~Sugamoto, Japan, 1979, KEK.

\bibitem{seesaw2} R.~N. Mohapatra and G.~Senjanovi{\'c}, \Journal
{\PRL}{44}{1980}{912}.

 
\bibitem{ellis} K. Benakli and J. Ellis, in preparation.

\bibitem{lept} R. Barbieri, J. Ellis and M.K. Gaillard, \Journal
{\PLB}{90}{249}{1980};\\ E.~K. Akhmedov, Z.~G. Berezhiani, and
G.~Senjanovi\'c, \Journal {\PRL} { 69}{1992}{3013}.



\bibitem{Smi} K. Benakli and A.Yu. Smirnov, \Journal {\PRL}{79}
{4314}{1997};\\ K. Benakli, hep-ph/9801303.

%%%%%%%%%%%%%%%%%%%%%%%%%%Axion%%%%%%%%%%%%%%%%%%%%%%%%%%%%%%%%
\bibitem{choi} K. Choi, {\em Phys. Rev.} {\bf D56} (1997) 6588;\\ 
T. Banks and M. Dine,\Journal {\NPB}{505}{445}{1997} .

%%%%%%%%%%%%%%%%%%%%%%%%%%correctedstuff %%%%%%%%%%%%%%%%%%%%%%%%

\bibitem{tye2} Z. Kakushadze and S.H. Tye, hep-th/9809147.


%%%%%%%%%%%%%%%%%%%%%%%%%%Pati-Salam %%%%%%%%%%%%%%%%%%%%%%%%





\bibitem{PS} J.~Pati and A.~Salam, \Journal {\PRD}  {10}{1974}{275}.





\end{thebibliography}
\end{document}